\documentclass[journal=jacsat,manuscript=article]{achemso}

\setkeys{acs}{keywords = true}
\usepackage[version=3]{mhchem} % Formula subscripts using \ce{}
\usepackage{comment}
\usepackage{graphicx}
\author{Mudasar Bashir}
\affiliation{International College of Semiconductor Technology, National Yang Ming Chiao Tung University, Hsinchu 30013, Taiwan}
\author{Andrew Sanchez}
\affiliation{International College of Semiconductor Technology, National Yang Ming Chiao Tung University, Hsinchu 30013, Taiwan}

\author{Pranaba Kishor Muduli}
\affiliation{Department of Physics, Indian Institute of Technology, Delhi, Hauz Khas, New Delhi 110016, India}

\author{Artur Useinov}
\email{artu@nycu.edu.tw}
\phone{+886 3 5712121 \#55910}
\affiliation[Unknown University]
{International College of Semiconductor Technology, National Yang Ming Chiao Tung University, Hsinchu 30013, Taiwan}
\title[An \textsf{achemso} demo]
  {Magnetoresistance of a point contact with a one-nanometer-wide constrained domain wall at different mean free path asymmetries.} 
\abbreviations{IR,NMR,UV}

\begin{document}

\begin{tocentry} 
\begin{minipage}{0.4\linewidth}
 \includegraphics[width=8.6cm]{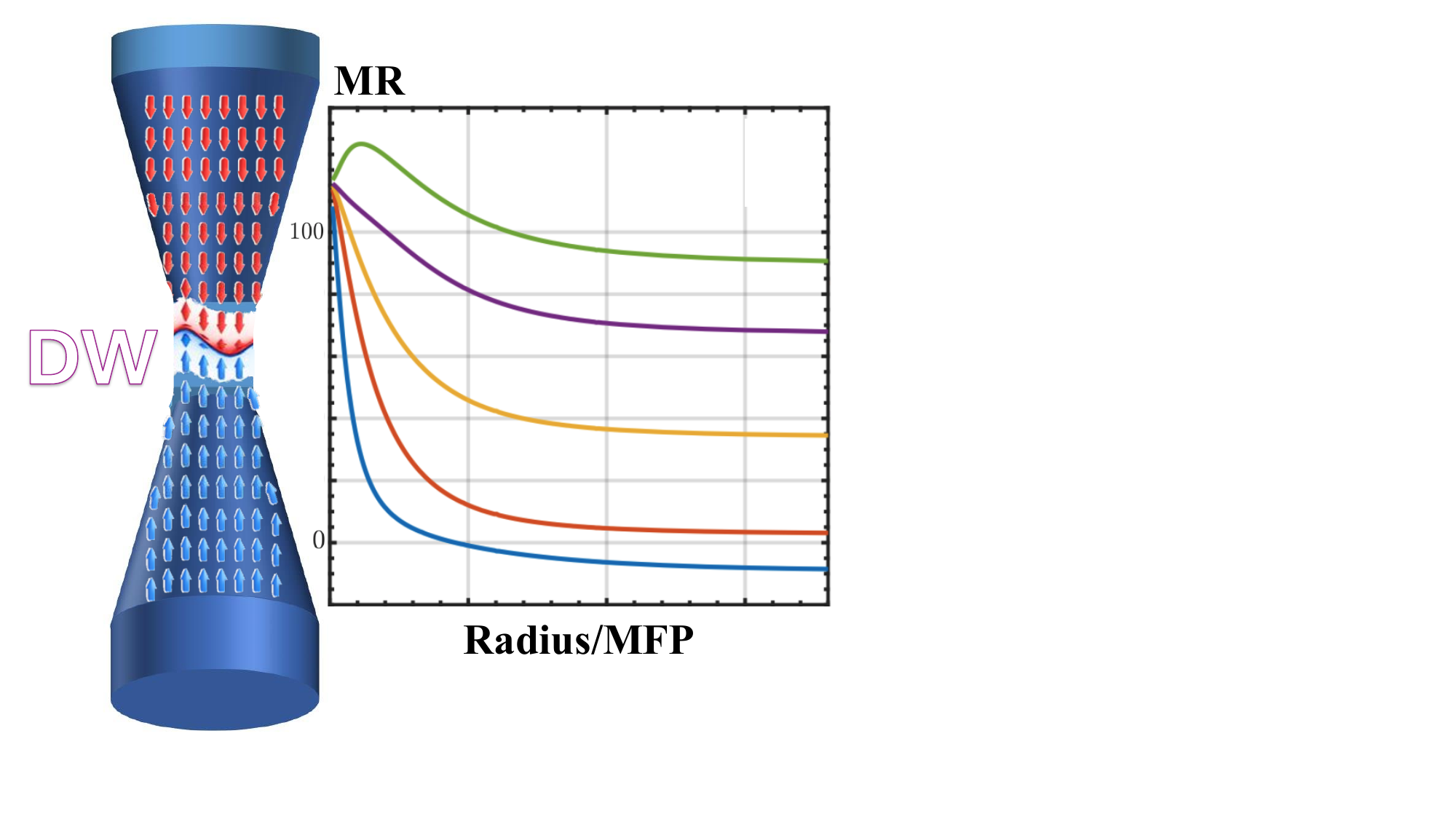}
 \end{minipage}
  \hfill
\begin{minipage}{0.35\textwidth}
 \textbf{Magneto-resis-tance ratio of a point-like contact with a 1~nm wide domain wall at different MFP asymmetries} 
 \end{minipage}
 \vfill

\end{tocentry}

\begin{abstract}
%%%%%%%%%%%%%%%%%%%%%%%%

This work presents a unified theoretical framework for spin-resolved electron transport in magnetic point contacts (PCs) in nanoscale dimensions. This work advances existing research by presenting a model which seamlessly transitions between Sharvin ballistic and Maxwell-Holm diffusive limits across the wide range of relevant contact sizes without incorporating empirical fitting factors. We analyzed the magnetoresistance (MR) of magnetic PCs formed with two ferromagnetic monodomains that may have parallel and antiparallel magnetization alignment, forming a constrained domain wall approximately 1.0~nm wide. The calculated MR exhibits strong dependence on scaling parameter (normalized contact radius), ratios of spin-dependent mean free paths, and Fermi wave-vectors. Furthermore, the calculated MR exhibits physically meaningful behavior over a wide range of spin-asymmetry parameters. In most regimes, the MR decreases with increasing normalized point-contact radius, becoming negative at some conditions. These results demonstrate that nanoscaled magnetic PCs have great efficiency in terms of magnetoresistnace change and promising for application due to their simplicity.

\begin{comment}
   This is an example document for the \textsf{achemso} document
  class, intended for submissions to the American Chemical Society
  for publication. The class is based on the standard \LaTeXe\
  \textsf{report} file, and does not seek to reproduce the appearance
  of a published paper.

  This is an abstract for the \textsf{achemso} document class
  demonstration document.  An abstract is only allowed for certain
  manuscript types.  The selection of \texttt{journal} and
  \texttt{manuscript} will determine if an abstract is valid.  If
  not, the class will issue an appropriate error.   
\end{comment}

\end{abstract}

\section{Introduction}
\label{sec1}

\begin{comment}
 This is a paragraph of text to fill the introduction of the
 demonstration file.  The demonstration file attempts to show the
 modifications of the standard \LaTeX\ macros that are implemented by
 the \textsf{achemso} class.  These are mainly concerned with content,
 as opposed to appearance.   
\end{comment}

Magnetic point-like junctions show great promise in the detection of single magnetic domain walls and skyrmions \cite{ArtSPIN22,UsAPL24}. It is critical to explore and understand the PC system over a wide range of parameters and transport conditions. Skyrmions and domain walls are key components in various designs for low-power spintronic race track memories, memristors, and logical chains \cite{LeqSciRep16,Luo21}. A common and analytically tractable model for a PC or nanocontact (NC) treats the contact as a circular aperture of radius $a$ having connection with two large reservoirs of electrons. The transport characteristics of the NC are conveniently expressed through the dimensionless ratio of $a$ to $l$ - the mean free path (MFP) of electron. This ratio, $a/l$, or its inverted value - the Knudsen number $K = l/a$, is usually obtained by fitting theoretical predictions to experimentally measured PC resistance \cite{Wex6}. When $l$ is determined from the material's resistivity measurement, the effective contact diameter can then be inferred from the fitted value of $K$. The model NC's diameter $d = 2a$ approximates the size of contact in cases where the detailed geometry is unknown. 

There are two key limiting regimes of electron transport through NCs. The first is the Maxwell-Holm, or diffusive, regime. This regime applies when the contact dimension greatly exceeds the MFP ($K \ll 1.0$) \cite{Max1,Hol3,Doudin}, and in this regime the conductance of a NC is $G_{M}$ which is calculated as
\begin{equation}
\label{e1}
G_{M} = \frac{2a}{\rho_{\rm V}},
\end{equation}
where $\rho_{\rm V}$ denotes the bulk resistivity. The latter may be expressed through the bulk conductivity $\sigma_{\rm V}$ of an isotropic metal as
\begin{equation}
\label{e2}
\rho_{\rm V}^{-1} = \sigma_{\rm V} = \frac{e^{2} n\, l}{\hbar k_{F}}
= \frac{e^{2} p_{F}^{2} l}{3\pi^{2} \hbar^{3}}
\end{equation}
in terms of $e$ (the electron charge), $k_{F}=p_{F}/\hbar$ (the Fermi wave-number), and $n = k_{F}^{3}/3\pi^{2}$ (the free-electron concentration). In this framework, where bulk MFP is $l = \hbar k_{F} \tau / m_{e}$, $m_{e}$ - the electron mass, it is governed by impurity scattering, lattice defects, electron–phonon, electron–electron interactions through the average time $\tau$ between collisions.

The second limiting case of electron transport is the Sharvin limit, or ballistic transport regime. This regime dominates when electrons traverse the constriction without undergoing scattering events \cite{Sha7}, i.e., for $K \gg 1$. In this situation, the Sharvin conductance ($G_S$) contains no dependence on $l$, reflecting the absence of any resistive processes along the electron trajectory:

\begin{equation}
\label{e3}
G_S  = \frac{{e^2 a^2 k_F^2 }}{{4\pi \hbar}} = G_0 N.
\end{equation}
The constant $G_0 = 2e^2 /h = 7.7481 \cdot \,10^{-5} \,\Omega ^{-1}$ is the quantum of conductance, and $N \simeq \left( {k_F a/2} \right)^2 $ is the whole amount of open channels \cite{Lesovik2011}.

One can readily derive relations expressing $G_{S}$, $\sigma_{\rm V}$, and $G_{M}$ in terms of one another: 
$G_{S} = \frac{3\pi a}{4K}\,\sigma_{\rm V}$, 
$\sigma_{\rm V} = \frac{4K}{3\pi a}\, G_{S}$, 
and 
$G_{M} = \frac{8K}{3\pi}\, G_{S}$. 
In his seminal work, Sharvin \cite{Sha7} obtained the asymptotic expression for the resistance: 
$R_{S} \simeq p_{F} / \left( e^{2} d^{2} n \right)$. 
Importantly, in past works the Sharvin conductance is often written as 
$G_{S} = 3\pi / \left( 16 R_{S} \right)$, 
up to a numerical factor $3\pi/16$ \cite{Tim4,Ert9,Gat11}. 
Further, the carrier density $n$ is a nontrivial function of $k_{F}$ in general case. Therefore, when the free-electron approximation is not applicable, $n$ should be corrected according to the electronic structure of the particular material. 
Both $n$ and $k_{F}$ can be determined from {\it ab-initio} computations.

There have been several attempts to establish models which effectively account for both the  ballistic (Sharvin) and diffusive (Maxwell) regimes in NCs. In 1999, Nikolić and Allen \cite{Nik20} revisited the Wexler formulation of the conduction for non-magnetic junctions through the orifice. They found electric potential in the presence of an electric field, solving static Boltzmann and Poisson equations in range of Fermi-Dirac statistics and Bloch-wave propagation approach. 
This treatment is frequently cited as one of the most accurate theoretical descriptions of the problem \cite{Tsymbal}. 
However, the form of their solution do not provide the smooth transition from Sharvin to Maxwell limits.

Concurrently, Mikrajuddin {\it et al.} \cite{Mik21} introduced an alternative resistance model based on solving the electrostatic Laplace equation, summing the resistances for infinitesimal shells which are bounded by equipotential surfaces inside the circular contact. A comparison between the Nikolić--Allen and Mikrajuddin {\it et al.} models reveals substantial discrepancies, further renewing interest in the underlying physics of the constriction problem and its improvement.
 
As the results of these works informed much of the emerging literature, the classical electrodynamic treatment of an orifice-type constriction increasingly produced a value for conductance composed of additive diffusive and ballistic contributions, accompanied by a nontrivial crossover regime in which residual terms persist. A more recent solution, derived from quasi-classical kinetic equations together with quantum boundary conditions, enables a description in which these residual terms vanish, thereby yielding a seamless transition between the classical limiting cases \cite{JMMM20, DiB20}. This work aims to leverage this improved solution to re-examine the characteristics of magnetic PCs with higher accuracy in relation to Ref.~\cite{EPJB07}, where similar MR simulations have been conducted using another approach.

\section{General model for the magnetic point contact in terms of spin-resolved transport}

\begin{figure}[!t]
\centering
\includegraphics[scale=0.40]{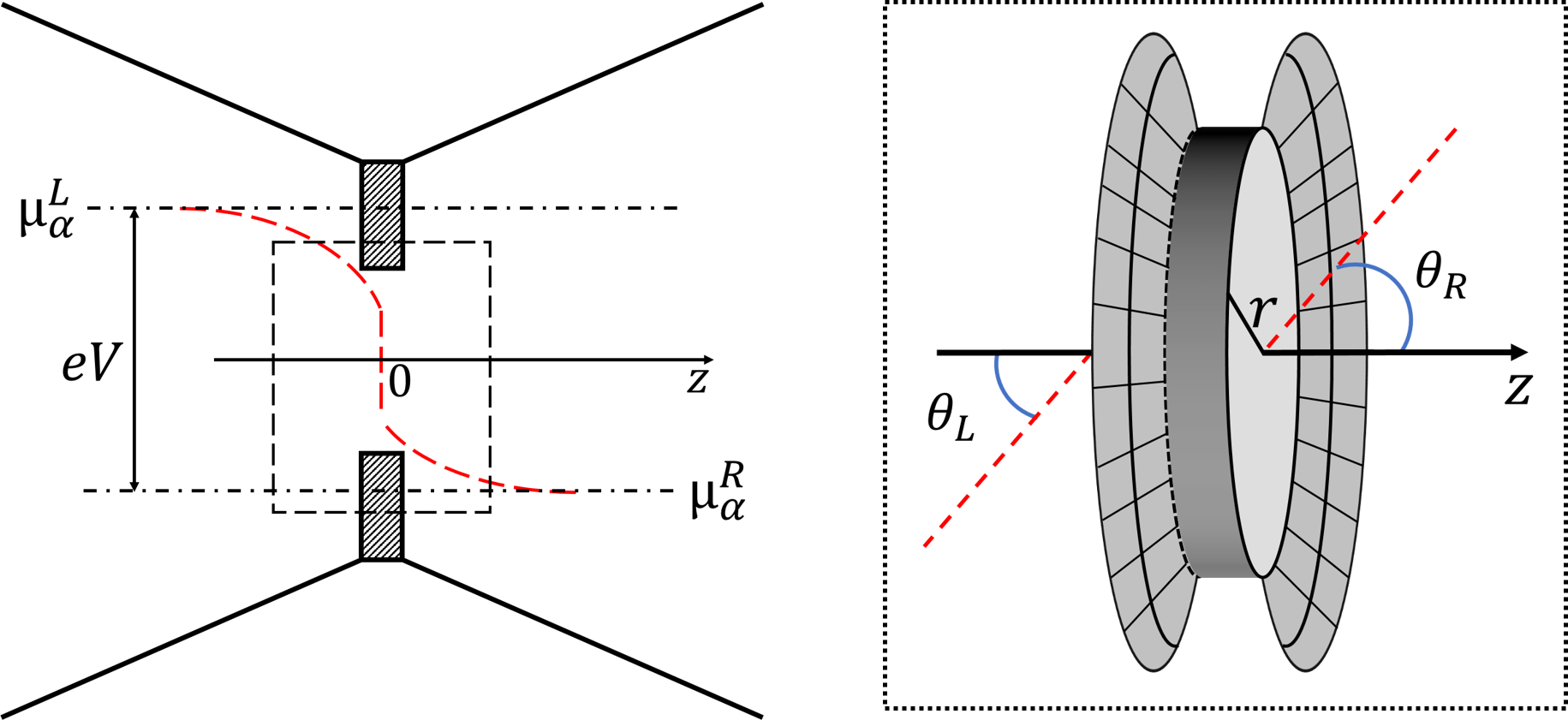}
\caption{The schematic view of a PC with a chemical potential drop, depicted by a dashed red curve. The contact area is shown within a dotted rectangle. An electron with an initial $p_{F,\alpha }^L$ and trajectory angle $\theta_{L,\alpha }$ transmits through orifice, resulting in a $p_{F,\alpha }^R$ and $\theta_{R,\alpha}$.} 
\label{fig1}
\end{figure}

A general model of a ferromagnetic (FM) hetero-contact is composed of two different FM metals, where the spin-dependent Fermi wave-numbers $k^c_{F,\alpha}$ on both sides ($c=L,R$) of the contact interface and the MFPs $l^c_{\alpha}$ ($\alpha = \uparrow, \downarrow$) are treated as tunable parameters. A circular conductive orifice of radius $a$, embedded in an otherwise non-conductive plane, separates the system into the left ($c=L$) and right ($c=R$) conductive half-spaces, and a magnetic domain wall is formed at the center of the constriction in the anti-parallel ($AP$) case of magnetization, Fig.\ref{fig1}.

The geometry matches the cylindrical coordinate system [${\bf{r}},\phi,z$] with the symmetry axis $z$. The applied voltage $V$ induces the electrical current $I^z  = I_ \downarrow ^z  + I_ \uparrow ^z$, being applied far away from the NC area. The solution for the charge current $I^z_\alpha$ with the spin projection $\alpha$ at positive $V$ applied to the right terminal, is characterized as follows \cite{DiB20}:

\begin{equation}
\label{e27}
I_\alpha ^{z}  = \frac{{e^2 ( k_{\rm{min}})^2 a^2 V}}{{2\pi \hbar}}\int\limits_0^\infty   dk\frac{{J_1^2 (ka)}}{k}F_{\alpha}(k),
\end{equation}
where $k$ is the radial variable in the contact plane originating from Fourier transform\cite{DiB20}, $k_{\rm{min}}$ is the minimum of $k_{F,\alpha}^{L}$ and $k_{F,\alpha}^{R}$, and ${{J}_{1}}\left(ka\right)$ is the Bessel function.

Despite existing similarity of expression (\ref{e27}) to past works \cite{Tag5,Use14,NUse2015}, the integrand function $F_{\alpha}$ is novelly defined as the following:
\small
\begin{equation}
\label{e27F}
F_\alpha  (k) = \mathop {\left\langle {x_L D_\alpha  } \right\rangle }\nolimits_{\theta _L }  - \left( {N_1\mathop {\left\langle {x_L W_L } \right\rangle }\nolimits_{\theta _L }  + N_2 \mathop {\left\langle {x_L W_R } \right\rangle }\nolimits_{\theta _L } } \right),
\end{equation}
\normalsize
where ${{D}_{\alpha}}$ is the quantum-mechanical transmission coefficient, $x_L  = \cos \left( {\theta _L } \right)$, and the angle between the $z$-axis and direction of the electron trajectory is ${{\theta }_{c,\alpha}}$, Fig.\ref{fig1}; the corner brackets $ \left\langle {...} \right\rangle _{\theta _L } $ is related with averaging of the content in spherical coordinate system [${\bf{k}},\theta,\varphi $]: $ {\left\langle {...} \right\rangle _{\theta _L }=\textstyle{1 \over {2\pi }}}\int\limits_0^{2\pi } {d\varphi } \int\limits_0^{\theta _{cr} } {\sin \left( {\theta _L } \right)\left( {...} \right)} {d \theta _L } = \int\limits_{\tilde x}^1 {(...) dx_L}$, where the limit $\tilde x=\cos \left({\theta_{cr}}\right)$ originates from the momentum conservation law of electrons along the NC plane. $\tilde x_{\alpha} = {\mathop{\rm Re}\nolimits} [\sqrt {(\delta_{\alpha}^{2}-1)/\delta_{\alpha}^{2}}]$ - lower integral limit for any $\delta_{\alpha}$, the value of $\delta$ is described by the following: $\delta_{\alpha}  = k_{F,\alpha }^L /k_{\alpha }^R(V)$, where $k_{\alpha }^R \left(V\right) = \sqrt { \left( {k_{F\alpha}^R } \right)^2  + \left( {2m_R e/\hbar ^2 } \right)V}$. The index $\alpha$, though not used in these relationships, appears in subsequent equations. Further key quantities and relationships are as follows:
\begin{equation}
N_1  = \left\{ {\left\langle {D_\alpha  } \right\rangle _{\theta _L } \left[ {2\left( {1 - \lambda _R } \right) + \lambda _2 } \right] - \left\langle {D_\alpha  } \right\rangle _{\theta _R }  \lambda _4 } \right\}\Delta ^{ - 1},
\end{equation}
 \begin{equation}
N_2  = \left\{ {\left\langle {D_\alpha  } \right\rangle _{\theta _R }  \left[ {2\left( {1 - \lambda _L } \right) + \lambda _1 } \right] - \left\langle {D_\alpha  } \right\rangle _{\theta _L } \lambda _3 } \right\}\Delta ^{ - 1},
\end{equation}
where
$\Delta = 4\left( {1 - \lambda _L } \right)\left( {1 - \lambda _R } \right) + 2\left[ { \lambda _1 \left( {1 - \lambda _R } \right) +  \lambda _2 \left( {1 - \lambda _L } \right)} \right] {-} \lambda _3  \lambda _4  +  \lambda _1  \lambda_2$,
and
\begin{equation}
\lambda_{c} = \frac{1}{1 + \left( k l^{c}_{\alpha} \right)^2}
\label{e8}
\end{equation}
\begin{equation}
\lambda_{1} = \left\langle
\frac{D_{\alpha}}
{\left( 1 + (k l^{L}_{\alpha})^2 \left( 1 - x_{L}^2 \right) \right)^{3/2}}
\right\rangle_{\theta_{L}}
\label{e9}
\end{equation}
\begin{equation}
\lambda_{2} = \left\langle
\frac{\delta \cdot x_{L} D_{\alpha}}
{\sqrt{x_{L}^2 + x_{cr}^2}
\left( 1 + (k l^{R}_{\alpha} \delta)^2 \left( 1 - x_{L}^2 \right) \right)^{3/2}}
\right\rangle_{\theta_{L}}
\label{e10}
\end{equation}
\begin{equation}
\lambda_{3} = \left\langle
\frac{\delta \cdot x_{L} D_{\alpha}}
{\sqrt{x_{L}^2 + x_{cr}^2}
\left( 1 + (k l^{L}_{\alpha})^2 \left( 1 - x_{L}^2 \right) \right)^{3/2}}
\right\rangle_{\theta_{L}}
\label{e11}
\end{equation}
\begin{equation}
\lambda_{4} = \left\langle
\frac{D_{\alpha}}
{\left( 1 + (k l^{R}_{\alpha} \delta)^2 \left( 1 - x_{L}^2 \right) \right)^{3/2}}
\right\rangle_{\theta_{L}}
\label{e12}
\end{equation}
\begin{equation}
\left\langle x_{L} W_{L} \right\rangle_{\theta_{L}}
=
\left\langle
\frac{x_{L} D_{\alpha}}
{\left( 1 + (k l^{L}_{\alpha})^2 \left( 1 - x_{L}^2 \right) \right)^{3/2}}
\right\rangle_{\theta_{L}}
\label{e13}
\end{equation}
\begin{equation}
\left\langle x_{L} W_{R} \right\rangle_{\theta_{L}}
=
\left\langle
\frac{x_{L} D_{\alpha}}
{\left( 1 + (k l^{R}_{\alpha} \delta)^2 \left( 1 - x_{L}^2 \right) \right)^{3/2}}
\right\rangle_{\theta_{L}}
\label{e14}
\end{equation}
\begin{equation}
\left\langle {D_\alpha} \right\rangle _{\theta _R } = \left\langle {\frac{{\delta  \cdot x_L D_\alpha  }}{{\sqrt {x_L^2  + x_{cr}^2 } }}} \right\rangle _{\theta _L }.
\label{e15}
\end{equation}
It should be noted that the derived set of equations, e.g. $\lambda_c$,  $\lambda_{1}$ to $\lambda_{4}$, as well as Eq.~(\ref{e13}) to Eq.~(\ref{e15}), are significantly improved in relation to the previous works \cite{Tag5, Use14, NUse2015}, while the tunnel-responsible, or ballistic term $\left\langle {x_L D_\alpha  } \right\rangle _{\theta _L }$ is the same. This difference originates from an updated solution to the integro-differential equation that more accurately account the solution for Green functions up to second-order by its derivatives \cite{JMMM20, DiB20}.

 The electron spin $\alpha$ is assumed to be conserved during transport through the NC, given the condition that the contact dimensions are smaller than the spin-diffusion length. Consequently, spin-flip scattering within the contact region is neglected. The symmetry of the system reveals the solution for the reversed bias $V$ with a negative (grounded) terminal on the right-hand side: $k_{F,\alpha }^L  \to k_\alpha ^R \left( V \right)$, $k_\alpha ^R (V) \to k_{F,\alpha }^L$. In the present work, we consider the approach of negligibly small positive voltage: the left side is assumed to be grounded, and the conduction band edge does not move with $V \to 0$, keeping the Fermi levels fixed. 
%If these conditions are not met, then the following will be true: $k_{\alpha}^R \left( V \right) = \sqrt { \left( {k_{F\alpha}^R} \right)^2   + \left( {2m_R e/\hbar ^2 } \right)V/2}$ and $k_{\alpha}^L \left( V \right) = \sqrt { \left( {k_{F\alpha}^L} \right)^2  - \left( {2m_L e/\hbar ^2 } \right)V/2}$, and $\delta  = k_{\alpha }^L(V) /k_{\alpha}^R (V) $. 
The transmission coefficient ${{D}_{\alpha}}$ mostly depends on the barrier for electrons within the NC area in the presence of the magnetic DW, which can be simulated in a range of linear potential energy profile~\cite{EPJB07}, while in the parallel ($P$) case of magnetization ${{D}_{\alpha}} = 1.0$. This is interesting given that the transmission ${{D}_{\alpha}}$ for 3D electron transport in bulk is originally characterized by the 1D potential energy profile $U(z)$ as a function of ${{\theta }_{c,\alpha}}$ and $V$, consisting of the $z$-axis projections of the Fermi wave-vectors $k_{\alpha }^{\bot L}   = k_{F\alpha}^{L} \cos (\theta _{L,\alpha })$ and $k_{\alpha }^{\bot R}   = k_{\alpha}^{R}(V) \cos (\theta _{R,\alpha })$ for 1D solution.

Symmetric and non-magnetic NC limit is approach which naturally coming from (\ref{e27}) assuming: $l^L =l^R = l$, $k_F^L = k_F^R = k_F$, $D_{\uparrow} = D_{\downarrow} = 1.0$ and thus $F_\downarrow(k) = F_\uparrow(k)$, $I_\uparrow^z = I_\downarrow^z$. By replacing the variable $y = ka$, one yields $\int_0^\infty  {dy\;J_1^2 (y)/y}  = 1/2$. The conductance $G = \frac{{dI}}{{dV}} =  \left( {I_\uparrow^z + I_\downarrow ^z }\right)/V$ for the small applied $V$ reads:
\begin{equation}
\label{e27G}
G  = 4 G _S \left( {\frac{1}{4} - \int\limits_0^\infty   \,\frac{{dy}}{y}\frac{{J_1^2 (y)}}{{1 + y^2K^2  + \sqrt {1 + y^2K^2 } }}} \right),
\end{equation}
which now cover both Maxwell-Holm and Sharvin limits exactly. Given this characteristic, one might expect that these equations would yield more precise $I$-$V$ curves for the non-magnetic nano-scale junctions, e.g. Pt-Pt, Au-Au, etc. 

\begin{figure*}[!t]
\centering
\includegraphics[scale=0.96]{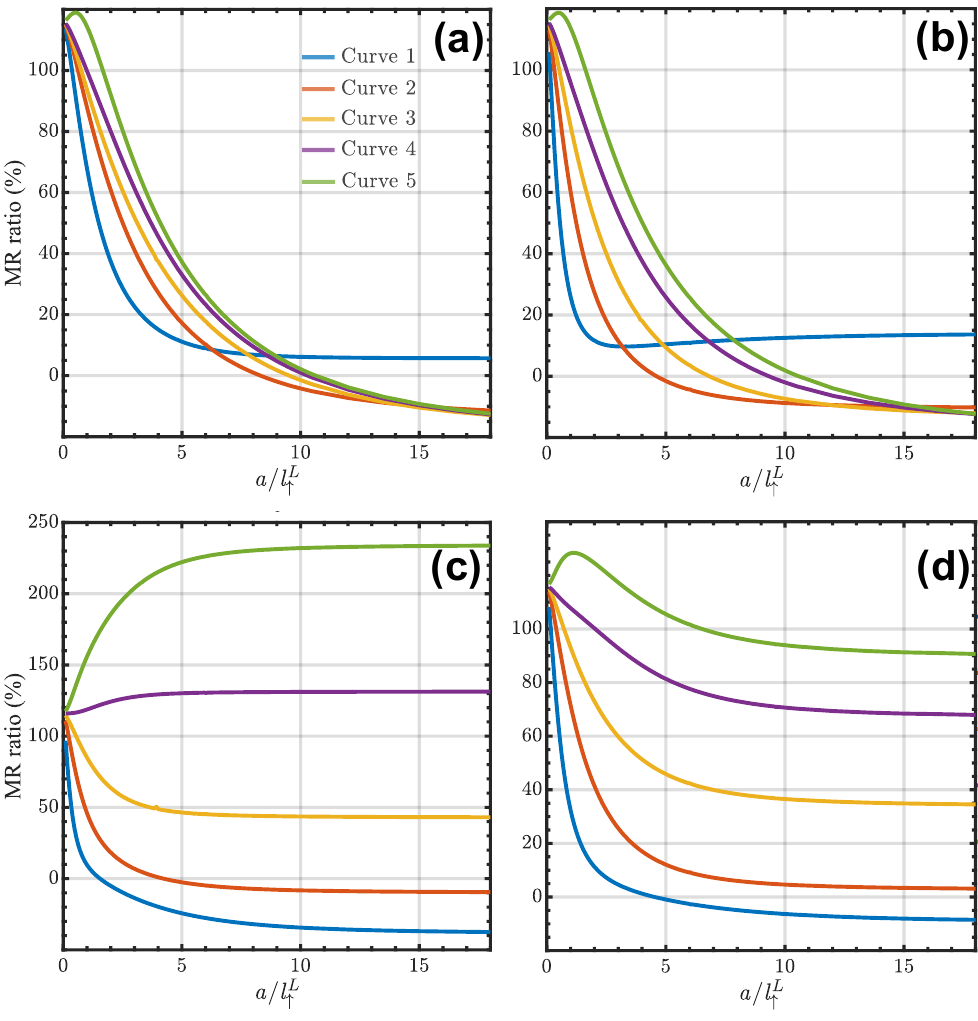}
\caption{Dependence of MR ratio on the normalized contact radius for different MFP asymmetry: a) $R_{ \downarrow }^{L} =l_{\downarrow}^{L}/l_{\uparrow}^{L} = 5.0$, $R_{R \downarrow }^{L} =l_{\downarrow}^{L}/l_{\downarrow}^{R}= 1.0;$ b) $R_{ \downarrow }^{L} = 2.0$, $R_{R \downarrow }^{L} = 1.0;$ c) $R_{ \downarrow }^{L}  = 1.0$, $R_{R \downarrow }^{L} = 1.0,$ and d) $R_{ \downarrow }^{L}  = 5.0$, $R_{R \downarrow }^{L} = 3.0$, where $R_{\downarrow }^{R} =l_{\downarrow}^{R}/l_{\uparrow}^{R}= 5.0, 2.0, 1.0, 0.5, \text{and}~0.2$~nm is set for curve 1 to curve 5, respectively, for each panel. For example, assuming $l_{\uparrow}^{L}=3.0~\text{nm}$, then $a=0.3$ - $54.0$~nm, and case (a): $l_{\downarrow}^{L}=l_{\downarrow}^{R}=15.0~\text{nm}$, $l_{\uparrow }^{R} =l_{\downarrow}^{R}/R_{\downarrow}^{R}= 3.0, 7.5, 15.0, 30.0, 75.0$~nm; case (b): $l_{\downarrow}^{L}=l_{\downarrow}^{R}=6.0~\text{nm}$, $l_{\uparrow }^{R} = 1.2, 3.0, 6.0, 12.0, 30.0$~nm; case (c): $l_{\downarrow}^{L}=l_{\downarrow}^{R}=3.0~\text{nm}$, $l_{\uparrow }^{R} = 0.6, 1.5, 3.0, 6.0, 15.0$~nm; and case (d): $l_{\downarrow}^{L}=15.0~\text{nm}$, $l_{\downarrow}^{R}=5.0~\text{nm}$, $l_{\uparrow }^{R} = 1.0, 2.5, 5.0, 10.0, 25.0$~nm for curve 1 to
curve 5, respectively.}
\label{fig2}
\end{figure*}

The complete analytical solution Eqs.~(\ref{e27})-(\ref{e27F}) is applied in the next section for magnetic nanocontacts. Two magnetic configurations are simulated: one parallel and one anti-parallel. This is done in the presence of spin up and spin down electron conduction channels at different spin asymmetries. In fact, for instance, a dramatic increase in the spin asymmetry of conduction-electron scattering may result due to the $3d$ and $4d$ solutes in iron \cite{dieny, jansen}. At low impurity concentrations, the band structure parameters of the contacting FM domains can remain unchanged while the spin-dependent MFP of each side can be varied independently over a wide range.
%The analytical solution Eqs.~(\ref{e27})-(\ref{e27G}) is applied in the next section for comparison with alternative theoretical approaches and experimental data available in literature. 
In the present work, some important parameters, responsible for MFP asymmetry, are defined as following: $R_{\alpha}^{L}=l_{\alpha}^{L}/l_{\uparrow}^{L}$, $R_{\alpha}^{R}=l_{\alpha}^{R}/l_{\uparrow}^{R}$, and $ R_{R \alpha}^{L} = l_{\alpha}^{L}/l_{\alpha}^{R}$, noticing that $R_{ \uparrow }^{L}=R_{ \uparrow }^{R}=1$. Combinations of these ratios (asymmetry coefficients) can be incorporated inside Eqs.~(\ref{e8})-(\ref{e14}), creating the ratios with one of MFPs. Here, $l_{\uparrow}^L$ is used for redefinition of existing $kl_{\alpha}^L$ and $kl_{\alpha}^R$ factors, standing inside $\lambda_{1}$,  $\lambda_{4}$ and other terms. At this case, making the dimensionless variables, it suggests the rescaling of the variable $k$ in Eq.~(\ref{e27}) into $\kappa = k l^L_{\uparrow}$. As a result, these factors are rescaled to $kl_{\alpha}^{L}\rightarrow kl_{\alpha}^{L}/\left( l_{\uparrow}^{L} \right) \times l_{\uparrow}^{L}$, or $kl_{\alpha}^{L}\rightarrow \kappa R_{\alpha}^{L}$, and, analogously, $kl_{\alpha}^{R}\rightarrow kl_{\alpha}^{R}/\left( l_{\uparrow}^{L} \right) \times l_{\uparrow}^{L}$, such that $kl_{\alpha}^{R}\rightarrow \kappa R_{\alpha}^{R}/R_{R\alpha}^{L}$.
Such substitution connects the bulk spin symmetry coefficients\cite{valet} $\beta=(1-\sigma_{\downarrow} / \sigma_{\uparrow})/(1+\sigma_{\downarrow} / \sigma_{\uparrow})$ with the defined asymmetry coefficients $R_{\downarrow}^{c} = l^c_{ \downarrow} / l^c_{\uparrow}$ as follows:
%\begin{equation}
%\label{e27beta1}
%\beta  = \frac{\rho_{\downarrow} - \rho_{\uparrow}} %{\rho_{\uparrow} +\rho_{\downarrow}} = \frac{1-\sigma_{\downarrow} / \sigma_{\uparrow}} {1+\sigma_{\downarrow} / \sigma_{\uparrow}},
%\end{equation}
\begin{equation}
\label{e27beta2}
\beta_{c}  = \frac{1-(\delta^{c})^{2} R_{\downarrow}^{c}} {1+(\delta^{c})^{2} R_{\downarrow}^{c}},
\end{equation}
since $\sigma_\alpha \propto k_{F \alpha}^{2} l_{\alpha}$.
In equation (\ref{e27beta2}), the spin polarization of the conduction band of $L$-($R$-) hand side FM metal is characterized by factor $\delta^{c} = k_{F \downarrow}^{c} / k_{F \uparrow}^{c}.$ Additionally, the contact's conductance depends on the ratio of the MFPs on the left and right sides, $R_{R\alpha}^{L}.$ If one MFP is known, e.g. $l_{L \uparrow},$ then the other three can be found using the $R_{\downarrow}^{L}, R_{\downarrow}^{R}$ and $R_{R \downarrow}^{L}$, accordingly. 

The following equation defines the magnetoresistance ratio in percentages:
\begin{equation}
\label{equMR}
\text{MR} = \frac{G^{P} - G^{AP}} {G^{AP}}\times 100\%,
\end{equation}
where the conductance for a parallel and anti-parallel configuration is defined as:  $G^{P(AP)}=G^{P(AP)}_{\uparrow}+G^{P(AP)}_{\downarrow}$. The conductance for spin-up (spin-down) channel $G^{P(AP)}_{\uparrow} \left(G^{P(AP)}_{\downarrow} \right)$ is defined as follows:  
\begin{equation}
\label{e27b}
G^{P(AP)}_{\alpha} = \frac{{e^2 {(k^{L}_{F,\alpha })}^2 \pi a^2}}{{\pi}^2} \int\limits_0^\infty d\kappa \frac{J_{1}^{2} (\kappa a/l^L_{ \uparrow})}{\kappa} F_{\alpha}(\kappa)^{P(AP)},
\end{equation}
where $a/l^L_{ \uparrow}$ is the dimensionless scaling variable, i.e., the normalized contact radius, which strongly influences MR value.

%%%%%%%%%%%%%%%%%%%%%%%%%%%%%

\section{Results and discussion}

The MR ratio is calculated for a variety of MFP and their spin asymmetries, Fig.\ref{fig2}. The rapid drops in MR ratio occur as the size of the contact increases beyond the MFP for either spin channel of conductance. According to the results, enhancing the MR ratio requires the conductance in range of the confined geometry to follow the ballistic, or quasi-ballistic regime. The results also imply that the reduction of MR with increasing contact cross-section is relatively weak. Furthermore, the limitations on achieving higher MR values by reducing the nanometric contact size, or approaching the ballistic limit are not restrictive for certain combinations of MFP asymmetries, as shown in Fig.\ref{fig2}c (curves 4, 5) and Fig.\ref{fig2}d (curve 5). Several factors contribute to the trend of the MR ratio, whether it is increasing, or decreasing. Increasing trends take place with condition when $l^R_\downarrow$ has a shorter value than others MFPs, inducing a significant reduction of $G^{AP}$. For example, the increasing MR ratio with saturation at $a/l^L_\uparrow \gg 1$ is observed for curve~4 and curve~5 at $l^R_\uparrow > l^R_\downarrow$, Fig.\ref{fig2}c, where $l^L_\downarrow/ l^L_\uparrow  = 1.0$, and  $l^R_\uparrow / l^R_\downarrow > 1.0$ are the supporting conditions for MR growth; a small bump is also observed in curve 5 of Fig.\ref{fig2}d, where $R^L_{R\downarrow} = 0.2$. This is because the conditions $l^R_\uparrow >> l^R_\downarrow$ and $l^L_\downarrow >> l^R_\downarrow$ result in an increase of the MR ratio while it remains in the ballistic regime (increasing the MR temporarily), before it starts to decrease as the diffusive regime becomes dominant with growing $a/l^L_\uparrow$. Similar effects take place for curve~5 in Fig.\ref{fig2}a and Fig.\ref{fig2}b. Several other conditions characterize a decreasing MR ratio, specifically the parameters $l^R_\downarrow = l^L_\downarrow$ with $R^L_\downarrow >1$ and $R^R_\downarrow < 5.0$. The presence of these condition leads to a deep and rapid decrease in MR values as a consequence of changing the sign with further increasing of scaling factor, e.g. the most visible in Fig.\ref{fig2}a, and Fig.\ref{fig2}b.

%\subsection{Outline}

% The document layout should follow the style of the journal concerned. Where appropriate, sections and subsections should be added in the normal way. If the class options are set correctly, warnings will be given if these should not be present.

\subsection{Conclusions}
In this study, we used a quasi-classical transport model to take a unified approach to compute spin-resolved conductance in magnetic point-like contacts. This theoretical framework successfully unifies spin-resolved ballistic, quasi-ballistic and diffusive transport conditions within unified analytical expression, avoiding any residual terms and empirical correction factors. The analytical solution obtained for the general spin-resolved case under cylindrical symmetry provides substantial practical usefulness. A key feature of the solution is that it incorporates a quantum-mechanical transmission coefficient for the linear potential profile of the barrier. This approach describes the potential energy change of conduction electrons passing the DW, and conserving the spin; and also the solution is combined with spin asymmetry coefficients enables the model to characterize asymmetric FM-based PCs. 

The results of this work were generated by applying the improved model, and they demonstrate that the MR ratio vs. scaling factor follows distinct trends. While increasing the scaling factor drives negative MR ratios in most cases, there are parameters of the system, where MR ratios increase above 100\%, having a saturation in the diffusive regime. And in other conditions the MR demonstrated a clear peak with further monotonic decreasing behavior. These behaviors are consistent across a range of spin asymmetry parameters, underscoring the model’s capability to describe spin-resolved transport in nanoscale magnetic junctions. The proposed model provides a solid basis for designing and analysing next-generation spintronic devices, such as sensors for detecting individual skyrmions, as well as magnetic Ni- or Co-based interconnects in nanoscale integrated circuits.

\begin{acknowledgement}

The authors thank Dara Mulia from National Yang-Ming Chiao-Tung University (ICST) for helping with this work and optimizing the code.  

\end{acknowledgement}

\begin{comment}
\begin{suppinfo}

This will usually read something like: ``Experimental procedures and
characterization data for all new compounds. The class will
automatically add a sentence pointing to the information on-line:

\end{suppinfo}    
\end{comment}

\bibliography{achemso-demo}

\end{document}